\relax
\documentclass[letterpaper]{article} 
\usepackage{aaai18}  
\usepackage{times}  
\usepackage{helvet}  
\usepackage{courier}  
\usepackage{url}  
\usepackage{graphicx}  
\usepackage{tabularx}
\usepackage{dcolumn}
\usepackage{booktabs}
\usepackage{url}
\usepackage{subfigure}
\usepackage{balance}  
\usepackage[usenames,dvipsnames,table]{xcolor}
\usepackage{enumitem}

\usepackage{balance}  
\usepackage{multicol}
\usepackage{multirow}
\usepackage{amssymb}	
\usepackage{amsmath} 

\usepackage[flushleft]{threeparttable}

\begin{document}

  \title{
    Political Discussions in Homogeneous and Cross-Cutting Communication Spaces\thanks{This paper has been supported by funding of the VolkswagenStiftung.}
    \\~\\ $[$\textsc{{Please cite the ICWSM'19 version of this paper}}$]$
    \\~\\
  }
\author{Jisun An \and Haewoon Kwak\\ Qatar Computing Research Institute \\ Hamad Bin Khalifa University \\ Doha, Qatar \\ \{jisun.an, haewoon\}@acm.org \And Oliver Posegga \\ University of Bamberg \\ Bamberg, Germany \\ oliver.posegga@uni-bamberg.de \And Andreas Jungherr \\ University of Konstanz \\ Konstanz, Germany \\ andreas.jungherr@gmail.com}

  \maketitle

\begin{abstract}
  
  Online platforms, such as Facebook, Twitter, and Reddit, provide users with a rich set of features for sharing and consuming political information, expressing political opinions, and exchanging potentially contrary political views. 
  In such activities, two types of communication spaces naturally emerge: those dominated by exchanges between politically homogeneous users and those that allow and encourage cross-cutting exchanges in politically heterogeneous groups. 
  While research on political talk in online environments abounds, we know surprisingly little about the potentially varying nature of discussions in politically homogeneous spaces as compared to cross-cutting communication spaces. 
  To fill this gap, we use Reddit to explore the nature of political discussions in homogeneous and cross-cutting communication spaces. In particular, we develop an analytical template to study \textit{interaction} and \textit{linguistic patterns} within and between politically homogeneous and heterogeneous communication spaces. 
  Our analyses reveal different behavioral patterns in homogeneous and cross-cutting communications spaces.  We discuss theoretical and practical implications in the context of research on political talk online. 
\end{abstract}

\section{Introduction}

  Internet platforms offer communication spaces in which users can express political opinions, share information, or exchange potentially contrary political views. In the early days, the internet's potential to foster political exchange across partisan political lines featured heavily in accounts of its political impact~\cite{Rheingold:1993aa}. 
  At present, discussions of the internet's perceived tendency to create politically homogeneous communication spaces, evocatively termed ``echo chambers'' or ``filter bubbles,'' dominate~\cite{Pariser:2011ly,Sunstein:2017aa}. 
  While there is the temptation to simplify the internet as either a space for homogeneous or cross-cutting political discussion and informational behavior, the empirical evidence regarding this is far from conclusive~\cite{Flaxman:2016aa,Webster:2014aa}.
  It is thus best to conceptualize the internet as a discussion space that allows for either homogeneous or cross-cutting political discussion and information exposure. But why does this matter?

  There is a rich debate in political science on the effects of the structure of political talk on democracy~\cite{Chambers:2003aa,Gamson:1992aa,mutz2006hearing}. 
  Some researchers expect that exchanges across partisan political lines are beneficial for healthy democracies~\cite{Barber:1984aa,Habermas:1989aa}. 
  In contrast, others point to the benefits of safe spaces of predominantly homogeneous communication environments in fostering political participation~\cite{Hibbing:2002aa,Mutz:2002aa} and the potential for the deterioration of political discourse in cross-cutting environments~\cite{Bail:2018aa,Theocharis:2016ac}. 
  In other words, while there is disagreement on what to expect from the structure of political talk across or along partisan lines, there is agreement on its importance in democracies.

  Accordingly, the nature of the internet and online platforms as political communication environments has been featured heavily in research. While much attention has been paid to the general conditions of political discourse online~\cite{Neuman:2011ij} and the structure of user interactions related to politics along or across partisan lines ~\cite{adamic2005political,conover2012partisan,an2014partisan}, we know little about the actual behavior of users in politically homogeneous or cross-cutting communication spaces.

  Social media services and other online platforms provide a constant stream of digital trace data~\cite{kane2014what} that can be used to gain insights into the nature of user behavior in the context of political discussions online~\cite{Jungherr:2015fj,Jungherr:2016ab}. 
  Digital trace data can be broken down into two different types~\cite{johnson2015emergence}: (1) traces of user activity, such as digital interactions between users; and (2) user-generated content.
  We study both data types in digital traces collected from \texttt{Reddit}\footnote{https://www.reddit.com} to identify specific patterns of political communication in homogeneous and cross-cutting information environments.
  
  First, through the analysis of the interaction structure, 
  we examine the degree to which users with different partisan leanings and preferences for homogeneous and cross-cutting communication spaces engage in political discussion in the respective environments. 
  Comparing user activity in subreddits with different political compositions, we classify users according to their behavior and investigate whether differences in the structure of political communication spaces also lead to differences in communication patterns among users.

  Second, through the analysis of user-generated content in the form of comments, 
  we examine linguistic patterns in political communication. This type of linguistic analysis provides the means to quantify the character of discussions and the relationship between speakers~\cite{Chambers:2003aa}. Thus, it is a valuable tool, which we use to study human behavior in largely politically homogeneous versus heterogeneous communication environments. 

  In combination, we identify whether interactions between users in largely homogeneous versus heterogeneous communication environments indeed differ as expected by democratic theory. 
  Normative theory places a high value on political discussion in heterogeneous political environments because it enables citizens to encounter arguments by supporters of opposing parties~\cite{Calhoun:1988aa,Habermas:1989aa} and to develop empathy for the other side~\cite{Benhabib:1992aa}. 
  
  If communication on Reddit conformed with these normative demands, we should expect to find users engaged in discussions with users of different political views. These exchanges should take the linguistic style of persuasive argument. This would, of course, presuppose that users actively engage in exchanges in politically cross-cutting communication spaces.
  As an alternative, Reddit users might be engaged mainly in largely homogeneous environments. Here, users should interact predominantly with others of the same viewpoint. These exchanges would likely take the linguistic styles of exchanging information, expressing opinions, and providing reinforcing statements. 
  This leads us to pose the following research questions:

  \begin{itemize}
    \item RQ1: To what degree do users with different political leaning interact with each other in politically cross-cutting environments?
    \item RQ2: Is there evidence that these users shift the linguistic style and word use of their posts between interactions in homogeneous versus cross-cutting environments?
  \end{itemize}

\section{Data Collection}

We study the nature of political exchanges in homogeneous and cross-cutting communication spaces. For this, we turn to Reddit, a prominent platform for political discussion and the exchange of political news~\cite{roozenbeek2017read}.

Reddit provides users with the choice of interacting in communication spaces that are politically largely homogenous or cross-cutting. They can create, subscribe, and contribute to self-organized, structured topical spaces--so called subreddits. 
Reddit users can post content to those subreddits and comment on content posted by others. Comments and posts can be voted ``up''- or ``down'' and scores are calculated based on the number of ``upvotes'' minus ``downvotes.''

To collect data from politically largely homogeneous communication spaces used during the 2016 U.S. presidential election, we focus on two subreddits: \texttt{/r/The\_Donald}\footnote{https://www.reddit.com/r/The\_Donald} for supporters of the leading Republican Party candidate Donald Trump, and \texttt{/r/HillaryClinton}\footnote{https://www.reddit.com/r/hillaryclinton} for supporters of the leading Democratic Party candidate Hillary Clinton.
As of September 15, 2018 both have 650,221 and 30,842 subscribers, respectively.\footnote{On the date of the US-election--November 8, 2016--the subscriber counts were reported as 273,677 and 35,002, respectively} Both subreddits have a clear policy of allowing each candidate's supporters to write posts while \textit{banning} users who post content critical of the respective candidate. Thus, we can reasonably assume that most of the users who leave comments in these subreddits are supporters of the respective candidate. 

To capture the data from politically heterogeneous communication spaces, we focus on  \texttt{/r/politics}\footnote{https://www.reddit.com/r/politics/} and \texttt{/r/news}.\footnote{https://www.reddit.com/r/news/} In both subreddits, users with different partisan leanings interact around general political topics and news items.
\texttt{/r/politics}, with 4,081,906 subscribers as of September 15, 2018, is self-described as ``the subreddit for current and explicitly political U.S. news.'' Users are only allowed to post articles, videos, or sound clips without any changes in the title of the original content. The content is restricted to articles dealing explicitly with US politics. 
Finally, \texttt{/r/news}, with 16,602,395 subscribers as of September 15, 2018, is a subreddit in which users share news items, and that explicitly excludes opinion pieces.

\begin{table}
  \small \frenchspacing
  \begin{center}
  \begin{tabular}{ccccc}
    \toprule
    space & subreddit & \# posts & \# comments & \# users \\
    \midrule
    $\circ$ & {/r/hillaryclinton} & 82,147 & 1,341,417 & 39,294  \\
    & {/r/The\_Donald}& 1,422,321& 11,635,535 & 298,464 \\
    \midrule
    $\dagger$ & {/r/politics} & 394,624 & 19,515,441 & 467,208 \\
    & {/r/news} & 754,708 & 7,393,738 & 51,694  \\
    \bottomrule
  \end{tabular}
      \begin{tablenotes}
        \small
        \item Note: We distinguish between homogeneous ($\circ$) and cross-cutting ($\dagger$) communication spaces.
      \end{tablenotes}
      \caption{Summary of our dataset \label{tab:dataset}} 
  \end{center}
\end{table}

For this study, we use the Reddit data collected by Jason Baumgartner and published at Pushshift.io,\footnote{http://files.pushshift.io/} which includes all publicly available submissions and comments from December 2005. 
We extract all posts and comments posted to the four subreddits between January 1, 2016 and December 31, 2016 from these data. This extends to approximately 2.5M posts and 39.8M comments. Table~\ref{tab:dataset} shows the basic statistics of our dataset.

\section{User and Interaction Types}

\subsection{User Classification by Political Leaning} 

To classify supporters of Clinton and Trump, we exploit their respective subreddits' exclusive content and banning policies and suggest the following identification rules. We define ``supporters'' of Clinton or Trump as active users who posted ten or more comments in \texttt{/r/hillaryclinton} or \texttt{/r/The\_Donald}, respectively. 
We then compute the average score (i.e., the number of upvotes minus downvotes) of each user's comments and exclude users whose average score is lower than zero in each subreddit. Negative average scores indicate that members of a specific subreddit reject those users. Thus, we do not consider them as ``supporters'' of the candidate of the corresponding community.\footnote{The initial score of a comment is +1.} 

Among 6,872 active users on \texttt{/r/hillaryclinton}, 321 users have a negative average score (i.e., more downvotes than upvotes). Among those users, 103 have appeared in \texttt{/r/The\_Donald}, and their average score on \texttt{/r/The\_Donald} is 7.477, which is quite high compared to the average score of 2.0 for all comments on \texttt{/r/The\_Donald}. This supports our assumption that such users are not ``supporters'' of the candidate in the corresponding community.
Similarly, among 67,034 active users on \texttt{/r/The\_Donald}, 430 users have a negative average score. Among them, 16 have appeared in \texttt{/r/hillaryclinton}, and the average score of their comments on \texttt{/r/hillaryclinton} is 2.717. 

Additionally, 530 users have non-negative average scores in both subreddits. We excluded them as well. In summation, we found 6,021 Clinton supporters and 66,074 Trump supporters. We denote Clinton supporter as $S_{C}$ and Trump supporter as $S_{T}$. 
Since we have no information to infer the political leaning of the remaining users, we refer to them as \textit{unassigned} users and denote as $S_{U}$.

To answer our first research question, we investigate the activity of each user group in the respective homogeneous and cross-cutting communication environments and compare their behavior.

\subsection{User Classification by Obtained Scores} \label{sec:user_classification}

Using users' obtained scores, we then differentiate between those who are and are not deemed helpful and informative by their communities.
We define users whose average comment score is higher than the 66th percentile of all users' scores in a given subreddit as ``high-scored users.'' We classify as ``low-scored users'' those whose average score is lower than the 33rd percentile of all users' scores. Users whose scores fall between these two percentile points are not considered.

The same user can be high-scored in one subreddit and low-scored in the other, meaning that the audiences of both subreddits perceive the value of that user's comments differently.
To examine robustly such differences between a pair of subreddits, we focus on four combinations of user groups based on score pairing in both subreddits: high-high, high-low, low-high, and low-low.
For the sake of argument, let us say that we compare the behavior of Clinton supporters in \texttt{/r/hillaryclinton} and \texttt{/r/politics}. A Clinton supporter with a high-high score pairing indicates that members of both subreddits, \texttt{/r/hillaryclinton} and \texttt{/r/politics}, find her comments helpful and informative. If her score pairing is high-low, it would indicate that members of \texttt{/r/hillaryclinton} find her comments laudable while members of \texttt{/r/politics} find her comments out of place.

Table~\ref{tab:high_low_comment} shows the number of users with each of the  score combinations in homogeneous and cross-cutting communication environments.

\begin{table}[h!]
  \begin{tabular}{cc|cc|cc}
    \toprule
     & & \multicolumn{2}{c|}{{/r/hillaryclinton}} & \multicolumn{2}{c}{{/r/The\_Donald}} \\
     & &  Low & High & Low & High \\
    \hline
    \multirow{2}{*}{{/r/politics}} & High & 182 & 1,958 & 142 & 1,053 \\
    & Low & 71 & 135 & 66 & 275 \\
    \hline
    \multirow{2}{*}{{/r/news}} & High & 1,380 & 11,647 & 1,239 & 11,394 \\
    & Low & 933 & 9,289 & 794 & 4,919 \\
    \bottomrule
  \end{tabular}
  \caption{Number of users based on their score pairings in homogeneous and cross-cutting communication environments}            \label{tab:high_low_comment}
\end{table}

\subsection{Interaction Type Classification}  
We further categorize the interaction patterns between the identified user types by political leaning.
We use the term ``interaction'' to indicate direct replies to contributions (i.e., posts and comments) of users. For example, when user A replies to user B's comment, we say that A initiates an interaction with B. As Reddit supports a hierarchical comment structure, who replies to whom can be easily identified.
We define three interaction types: 1) a supporter of one candidate communicates with another supporter of the same candidate; 2) a supporter communicates with a supporter of the opposing candidate; and 3) a supporter communicates with an unassigned user. 
Considering whether a Trump or a Clinton supporter initiates the interaction, we can have 2 $\times$ 3 = 6 combinations of different interaction types. 
For example, we note the interactions from Clinton supporters to Clinton supporters ($S_{C} \rightarrow S_{C}$), from Clinton supporters to Trump supporters ($S_{C} \rightarrow S_{T}$), and from Clinton supporters to unassigned users ($S_{C} \rightarrow S_{U}$), and note the same three for interaction types from Trump supporters.

\section{Methodology}
\subsection{Analysis of Interaction patterns}
\label{sec:method_zcore}
We first examine the degree to which supporters of each candidate engage in cross-cutting conversations. Considering the strongly divergent numbers between users of each type, we build a null model to estimate the expected frequency of each interaction type and compare it with the observed frequency from our data.
The null model is a set of randomly generated ensembles that shuffle authors of posts and comments while preserving the original structure of posts and comments from the actual data. In other words, we can estimate the expected frequency of each interaction type occurring at random.
The idea behind this null model is similar to randomly rewiring links while preserving the degree distribution of complex networks when the null model of a network is required~\cite{maslov2002specificity}.
We can quantify the cross-cutting interaction behavior and test its statistical significance by computing the \textit{z}-scores and comparing the number of interaction types in the actual data with that in the null model.

Let us give an example of interactions initiated by Clinton supporters ($S_{C}$) with Trump supporters ($S_{T}$). We denote by $|N_{S_{C} \rightarrow S_{T}}|$ the number of interactions from $S_{C}$ to $S_{T}$ observed in a given subreddit. 
To construct a null model, we create 1,000 randomly generated ensembles; each has the same number of comments, the same structure of posts and comments, and the same set of authors with the original data, but has randomly shuffled authors of the posts and comments.  
We then count how many ${S_{C} \rightarrow S_{T}}$ interactions are observed in each ensemble and denote the count as $|R_{S_{C} \rightarrow S_{T}}|$. We compute the average and the standard deviation of $|R_{S_{C} \rightarrow S_{T}}|$ across the 1,000 ensembles. The corresponding \textit{z}-score $Z({S_{C} \rightarrow S_{T}})$ is then computed as follows:
\begin{equation}
    Z({S_{C} {\rightarrow} S_{T}})=\frac{|N_{S_{C} \rightarrow S_{T}}|-avg(|R_{S_{C} \rightarrow S_{T}}|)}{std(|R_{S_{C} \rightarrow S_{T}}|)}
\end{equation}

A higher \textit{z}-score indicates that the corresponding interaction type is likely to occur in the actual dataset more frequently than by chance. Thus, by calculating the \textit{z}-score for each interaction type, we examine the tendency of various interaction types to emerge in homogeneous and cross-cutting communication spaces.

\subsection{Analysis of Linguistic Patterns}
\label{sec:linguistic_features}

Our second research question focuses on the linguistic patterns used by the supporters of each candidate within homogeneous and cross-cutting environments. We analyze comments left by the supporters of the two candidates from three perspectives: 1) their linguistic style; 2) the similarity of their vocabulary; and 3) the semantic difference in their words.

\subsubsection{Linguistic style} 

We compare users' linguistic styles captured from comments in homogeneous and cross-cutting spaces. 
To study the differences in linguistic style systematically, we use the Linguistic Inquiry and Word Count (LIWC)~\cite{pennebaker2001linguistic}. 
First introduced in 2001~\cite{pennebaker2001linguistic}, LIWC has been used widely in the analysis of textual content in various social media across multiple domains~\cite{tausczik2010psychological}. 
It has also been used to analyze political communication, in particular to measure differences between Democrats and Republicans in the United States~\cite{sylwester2015twitter,preoctiuc2017beyond}. 

The 2015 version of LIWC automatically counts word frequencies for 93 categories. The internal dictionary has been constructed manually to reflect psychological theory. This includes the identification of parts of speech, topical categories, and emotions. LIWC measures the length-normalized value for each of the categories in a given text. 
We compare each of the LIWC categories in interactions between supporters of the same candidate (e.g., $S_{C} \rightarrow S_{C}$) with interactions between supporters of opposing candidates (e.g., $S_{C} \rightarrow S_{T}$) by conducting the paired $t$-test with Bonferroni correction for multiple comparisons.

In addition to the LIWC analysis, we also identify \textit{persuasive} linguistic style proposed by Tan et al.~\shortcite{tan2016winning}.
They find that, in \texttt{/r/ChangeMyView}, persuasive arguments tend to have a greater number of words (denoted as \# words); more personal pronouns (i.e., first-person singular pronouns (1SG), first-person plural pronouns (1PL), and second-person pronouns (2)); fewer positive words (pos.); more negative words (neg.); more question marks (?); more quotations (quot.); more calm words, indicating lower level of arousal (arousal); and a lower valence level (valence). Although this finding has not been validated with other data, it is reasonable to assume that similar characteristics can be shared within the users of the same service, Reddit.

As arousal and valence are not LIWC categories, we follow Tan et al. (\citeyear{tan2016winning}) and use word/score pairs provided by~\cite{brysbaert2014concreteness,warriner2013norms} to calculate the average arousal/valence score for each comment.

In our setting, we can expect users to engage in a more persuasive linguistic style when communicating with the supporters of the opposing candidate than with those of the same candidate, given that perceive and use Reddit as an online space for constructive political discussion.

\subsubsection{Similarity of vocabulary} 

Next, we use Jaccard similarity to measure the similarity of vocabulary used by supporters across the interaction types. 
For each of the interaction types, we combine all respective comments in cross-cutting communication spaces (i.e., \texttt{/r/politics} and \texttt{/r/news}). 
We then use a simple but effective method, Term Frequency -- Inverse Document Frequency (TF-IDF), to find important words in each interaction type. We consider each comment as a document for TF-IDF computation. 
TF-IDF value increases by a high term frequency in a comment and a low document frequency of the term in the corpus. Thus, it filters out common, less significant words.
For each interaction type, we rank words by summing the TF-IDF values across all comments, extract the top words in the ranking, and compute the Jaccard similarity across the interaction types. The Jaccard similarity is defined as:
  $Jaccard(A,B) = \frac{A \cap B}{A \cup B}$
\noindent where $A$ and $B$ are the sets of the top $k$ words by TF-IDF values of the two different interaction types. 
For example, when $k$ equals 10, we compare the top 10 words by TF-IDF values among Clinton supporters who talk to other Clinton supporters to those who talk to Trump supporters.

\subsubsection{Semantic differences} 
The similarity of vocabulary shows the similarity of the word choice. However, it is possible that the same word can be used differently. 
The semantic difference of the same words in different corpora has been actively studied in recent years~\cite{an2018semaxis,hamilton2016diachronic}.
We measure semantic differences of words by using word embeddings as proposed by Hamilton et al.~\shortcite{hamilton2016diachronic}. 
The entire procedure can be summarized in three steps:

\textit{1) Training word embeddings:} We train our model to learn word representation in a vector space. We use the skip-gram model~\cite{mikolov2013efficient}, which predicts the context surrounding a given word. 
After a sufficient degree of training, the model learns a deep representation of each word that is predictive of its context. 
We can then use these representations, called neural word embeddings, to map words onto a vector space. 
Our parameters for learning are: size 300; a symmetric context window of size five; a minimum word count of ten; negative sampling; and down-sampling of frequent terms as suggested by Levy et al. (\citeyear{levy2015improving}). 
We construct word embeddings based on the corpus of each  interaction type, resulting in having word vectors for all words in the corpus.

\textit{2) Aligning embeddings:} To compare word vectors in different embeddings, we must ensure that the word embeddings are aligned to the same coordinate axes because word embeddings are constructed in a  stochastic way.  We align the learned word embeddings from  different interaction types by using orthogonal Procrustes as proposed in \cite{hamilton2016diachronic}. 

\textit{3) Quantifying semantic differences:} Once we have aligned word embeddings of different interaction types, we can compute directly the cosine distance of word's vector representations between different word embeddings. The resulting cosine distance measures semantic differences of the word between the corresponding interaction types.

\section{User Activity and Interactions in Politically Homogeneous and Cross-cutting Communication Spaces}

In this section, we answer our first research question: \textit{To what degree do users with different political leanings interact with each other in politically cross-cutting  environments?}

\subsection{Level of User Activity in Both Spaces}

We first examine patterns in user activity. 
We find that 6,021 Clinton supporters and 66,074 Trump supporters are very active with an average of 165 comments in \texttt{/r/hillaryclinton} and 144 in \texttt{/r/The\_Donald}, respectively. 
Irrespective of the differences in the total number of active supporters in both subreddits, individuals' activity levels appear to be rather similar, with supporters of Hillary Clinton being slightly more active. 
Also, we find that, in both subreddits, supporters with a greater number of comments tend to have a higher average score of their comments: the Spearman's ranking correlation coefficient between the number of comments left by a user and the average score of their comments is 0.269 ($p$ $<$ $0.005$) for \texttt{/r/hillaryclinton} and 0.322 ($p$ $<$ $0.005$) for \texttt{/r/The\_Donald}. 

\begin{table}[th!]
\footnotesize \frenchspacing
\begin{center}
  \begin{tabular}{cccc}
    \toprule
     & & {Clinton} & {Trump}  \\
     & & {Supporters} & {Supporters} \\
     \midrule
    \multirow{3}{*}{\textbf{Users}} & supporters & 6,021  & 66,074 \\
     & {/r/politics} & 4,689(78\%) & 40,236(61\%)\\
    & {/r/news}& 2,922(49\%) & 32,083(49\%)\\
    \midrule
    \multirow{3}{*}{\textbf{Comments}} &{/r/hillaryclinton} & 445,945 & -\\
     &{/r/The\_Donald} & - & 5,293,143\\
     &{/r/politics} & 1,403,818 & 3,372,726 \\
    &{/r/news}& 90,115 & 902,473\\
    \midrule
    \multirow{3}{*}{\textbf{Submissions}}     & {/r/hillaryclinton} & 42,570 & -\\
      &{/r/The\_Donald} & - & 710,149\\
     &{/r/politics} & 87,574 & 141,523\\
    &{/r/news} & 5,786 & 32,665\\    
    \bottomrule
  \end{tabular}
    \begin{tablenotes}
      \small
      \item Note: ``Submissions'' refer to the number of posts in which users leave at least one comment.
    \end{tablenotes}
    \caption{Contributions by active users in homogeneous and cross-cutting  communication environments}
    \label{tab:crossing_users}
    \end{center}
\end{table}

Table~\ref{tab:crossing_users} shows that significant shares of Clinton supporters and Trump supporters are active in cross-cutting spaces (i.e.,  \texttt{/r/politics} and \texttt{/r/news}).  
In \texttt{/r/politics}, 77.9\% of Clinton supporters are also active, contributing on average 299 comments. In total, they left 1.4M comments in \texttt{/r/politics}. 
Trump supporters are also active in \texttt{/r/politics} but to a lesser extent than Clinton supporters. 
Around 60.9\% of Trump supporters post an average of 84 comments each in \texttt{/r/politics}. Given their overall stronger numbers, this runs at a total of 3.4M comments. 
This points to a first interesting difference between supporters of both candidates. Although Trump supporters dwarf Clinton supporters in raw numbers both in homogeneous and cross-cutting communication environments, they appear to be more comfortable interacting among like-minded users than are Clinton supporters. 
Even if they expose themselves to cross-cutting political talk, they do so much less actively than Clinton supporters. 

In \texttt{/r/news}, both supporter groups show much lower activity compared to those in \texttt{/r/politics}. This might be due to the policy of \texttt{/r/news}, predominantly pointing contributions on electoral politics to the subreddit \texttt{/r/politics}.

\subsection{User Interactions in Both Spaces}

We then move on to examine how supporters of both candidates interact in cross-cutting communication spaces. 
We count how many posts in cross-cutting spaces have comments from supporters of either candidate. In \texttt{/r/politics}, there are 172,989 posts in which supporters of either candidate left comments. In 66,469 (38.4\%) of them, supporters of both candidates left comments. 
In \texttt{/r/news}, we find a lower proportion of posts in which supporters of both candidates left comments. Of 33,163 submissions that have comments from supporters of either candidate, 5,330 (16.1\%) posts have comments from the supporters of both candidates. 
We can thus conclude that most comment threads are exclusively the domain of supporters of one or the other candidate. 
The supporters of both candidates, however, actively left comments on a considerable number of posts and thus are likely to be exposed to comments by supporters of the opposing candidate. 
So, do supporters of the opposing candidates actually interact by comments?

To measure whether supporters of opposing candidates interacted more frequently than chance would suggest, we calculate \textit{z}-score by comparing the number of actual interactions with those simulated in a  null model (see Section~\ref{sec:method_zcore}).
Higher \textit{z}-score points to the corresponding interaction between users types being observed more often than chance would indicate. 
Table~\ref{tab:ninteractions_zscore} reports the number of occurrences of different interaction types with their \texttt{z}-scores.
We note that all our methods are designed to handle the differences of the activities of the two supporter groups and all the group differences reported in this study are statistically significant.

\begin{table}[th!]
  \begin{center}
    \footnotesize \frenchspacing
    \begin{tabular}{@{}c@{}@{}r@{}@{}r@{}@{}r@{}@{}r@{}}
      \toprule
      \multicolumn{1}{c}{Type} & \multicolumn{1}{c}{{politics}} & \multicolumn{1}{c}{{news}} & \multicolumn{1}{c}{{clinton}} & \multicolumn{1}{c}{{donald}} \\
      \midrule
      \textbf{$S_{C} \rightarrow S_{C}$} & 114.4K (16.0) & 1.5K (16.5) & 754.5K (10.6) & 37.0 (32.4)\\
      \textbf{$S_{C} \rightarrow S_{T}$} & 268.4K (29.3) & 14.5K (57.9) & 11.4K (-3.7) & 3.8K (9.3)\\
      \midrule
      
      \textbf{$S_{T} \rightarrow S_{T}$} & 479.6K (-16.5) & 109.6K (26.9) & 490.0 (5.5) & 7.9M (33.4)\\
      \textbf{$S_{T} \rightarrow S_{C}$} & 295.3K (24.3) & 16.4K (37.1) & 12.7K (8.9) & 3.8K (0.9)\\
      \bottomrule
    \end{tabular}
    \caption{Number of occurrences and its \texttt{z}-score by interaction type.}
    \label{tab:ninteractions_zscore}
  \end{center}
\end{table}

The table shows that interaction associated with Trump and Clinton supporters rather than unassigned users have high \textit{z}-scores, meaning they are empirically observed more often than in the random model. 
This is driven largely by supporters of either candidate who are less likely to interact with unassigned users than chance but more likely to interact with other users supporting either candidate.
The interactions between supporters of either candidate in \texttt{/r/politics} and \texttt{/r/news} show that users not only participate in parallel but actually do interact across party lines (\textit{z}-scores are 29.3 and 24.3 for $S_{C} \rightarrow S_{T}$ and $S_{T} \rightarrow S_{C}$, respectively, in \texttt{/r/politics} and 57.9 and 37.1 in \texttt{/r/news}). 
Here, interactions across party lines even surpass interactions among supporters of the same candidate (\textit{z}-scores are 16.0 and -16.5 for $S_{C} \rightarrow S_{C}$ and $S_{T} \rightarrow S_{T}$, respectively, in \texttt{/r/politics} and 16.5 and 26.9 in \texttt{/r/news}). This demonstrates that the two subreddits, \texttt{/r/politics} and \texttt{/r/news}, served as cross-cutting communication spaces in which supporters of the different candidates met and interacted. 

These cross-cutting interactions even appear, although to a lesser degree, in the predominantly homogeneous candidate subreddits.
While Clinton supporters are less likely to comment on Trump supporters' comments in \texttt{/r/hillaryclinton}, there exists significant interactions from Trump supporters to Clinton supporters in \texttt{/r/hillaryclinton}. Also, as indicated by positive \textit{z}-scores, Trump supporters do comment on Clinton supporters' comments in \texttt{/r/The\_Donald}.

The prevalent interactions between supporters of the two candidates show that online platforms do not necessarily force users into politically homogeneous ``echo chambers''~\cite{Sunstein:2017aa}. 
Rather, users actively seek others with diverse political opinions and exchange comments with them. Yet evidence of exchanges alone tells us little about their quality. For this, we move on to the linguistic analysis of the content of these exchanges. We focus in particular on how supporters' linguistic styles shift in politically homogeneous and cross-cutting communication spaces.


\section{Linguistic Styles, Vocabulary, and Semantics in Politically Homogeneous and Cross-cutting Communication Spaces}

In this section, we answer our second research question: \textit{Is there evidence that users shift the linguistic style and word use of their posts between interactions in homogeneous versus cross-cutting environments?}

\subsection{Linguistic Styles in Homogeneous and Cross-cutting Communication Spaces}

We have shown that supporters of the two candidates actually interact in cross-cutting communication environments. But what is the quality and type of these interactions? 
As interactions can take the form either of carefully worded exchanges about politics or strongly worded insults, measuring their quality and type is essential to understanding the nature of these interactions.
We focus in particular on their \textit{linguistic styles}. 
Characterizing linguistic styles allows us to assess  whether:
1) supporters shift their linguistic styles when interacting in homogeneous and heterogeneous communication spaces; 
2) supporters shift their linguistic styles when interacting with the supporters of the same candidate compared to when interacting with supporters of the opposing candidate; and 
3) interactions between supporters of the opposing candidate tend to follow a persuasive style or whether they were more expressive in nature.\footnote{We note that all results we report below are statistically significant at $p<$0.005. Due to space limitations, we focus here on discussing the most relevant findings. Omitted results are consistent with the reported findings and our conclusions.}

\subsubsection{Are supporters shifting their linguistic styles when communicating in homogeneous spaces compared to cross-cutting spaces?}

\begin{table*}[t!]
  \frenchspacing
  \begin{tabular}{p{8.3cm}p{0.1cm}p{8.3cm}}
    \toprule
    Dominant LIWC categories in \textbf{homogeneous} spaces & & Dominant LIWC categories in \textbf{cross-cutting} spaces \\
    \midrule
    \#words, f.ppron, f.1SG, f.percept, f.affiliation, f.reward, f.future, f.exclam, f.time, tone, f.we, f.see, f.feel, f.relative, authentic & & f.2, f.negate, f.cause, f.differ, f.work, f.money, f.qmark, f.period, clout, f.interrog, f.social, f.cogproc, f.certain, f.power, f.risk, f.death \\
    \bottomrule
  \end{tabular}
  \caption{LIWC categories that changed significantly in interactions between supporters of the same candidate in homogeneous spaces and  cross-cutting spaces.  Features passed a Bonferroni-corrected significance test.}
  \vspace{-1mm}
  \label{tab:liwc_across_subreddits}
\end{table*}
 
We begin by exploring the linguistic patterns of interactions between supporters of the same candidate (e.g., $S_{C} \rightarrow S_{C}$) in homogeneous (e.g., \texttt{/r/hillaryclinton}) and  cross-cutting communication spaces (e.g., \texttt{/r/politics} or \texttt{/r/news}). 
We have four interaction types to consider: 1) $S_{C} \rightarrow S_{C}$ in  \texttt{/r/hillaryclinton} vs. \texttt{/r/politics}, 2) $S_{C} \rightarrow S_{C}$ in  \texttt{/r/hillaryclinton} vs.  \texttt{/r/news}, 
3) $S_{T} \rightarrow S_{T}$ in  \texttt{/r/The\_Donald} vs.  \texttt{/r/politics}, and 4) $S_{T} \rightarrow S_{T}$ in  \texttt{/r/The\_Donald} vs. \texttt{/r/news}. 
Table~\ref{tab:liwc_across_subreddits} reports LIWC categories that are consistently used more often either in homogeneous or cross-cutting spaces.

In general, supporters appear more open, happy, and comfortable in politically homogeneous environments than when they interact in cross-cutting communication spaces; they use more words, their tone is more positive, and their words more often refer to perceptual processes (i.e., seeing and feeling).
In homogeneous environments, supporters express greater group responsibility by using more third-person plural pronouns (``we''). This can be explained by a shared social identity between supporters and a subjective sense of belonging~\cite{tajfel1979integrative}. Social identity theory also explains group members' desire to evaluate their groups positively~\cite{turner1987rediscovering,kwak2015exploring}, which could manifest in happy, comfortable, and positive tone style in their language use. 
We also find that supporters use more affiliation and reward words, indicating that they talk more about their determination and ambition in homogeneous spaces as compared to cross-cutting spaces. 

Conversely, we observe a greater use of second-person pronouns (``you'') in cross-cutting spaces. This outgroup behavior can be seen as the mirror image of the ingroup behavior observed in homogeneous spaces~\cite{turner1987rediscovering}. A greater use of words pointing to cognitive processing, causal reasoning, and negations indicates that the supporters use more complex language and sophisticated reasoning in cross-cutting spaces. Also, a higher use of question marks implies that questions are more common in political contributions to cross-cutting spaces.

\subsubsection{Do supporters change linguistic style when talking to supporters of opposing candidates?}

\begin{table}[t!]
  \footnotesize \frenchspacing
  \begin{tabular}{ccccccc}
    \toprule
    & \multicolumn{3}{c}{\tiny{$S_{C} \rightarrow S_{C}$ vs $S_{C} \rightarrow S_{T}$}}  & \multicolumn{3}{c}{\tiny{${S_{T} \rightarrow S_{T}}$ vs ${S_{T} \rightarrow S_{C}}$}} \\
    \hline
    & clinton & politics & news & donald & politics & news  \\
    \hline
    \tiny{$>>>$} & 12 & 2 & 0 & 12 & 7 & 5 \\
    \tiny{$>>$} & 2 & 2 & 0 & 2 & 0& 1 \\
    \tiny{$>$} & 1 & 2 & 0 & 0 & 4 & 2 \\
    \hline
    \tiny{$<$} & 2 & 0 & 0 & 4 & 1 & 3 \\
    \tiny{$<<$} & 0 & 1 & 1 & 1 & 4 & 0 \\
    \tiny{$<<<$} & 9 & 4 & 1 & 16 & 5 & 0 \\
    \midrule
    Total & 26 & 11 & 2 & 35 & 21 & 11 \\
    \bottomrule
      \end{tabular}
  \caption{Number of LIWC categories that shift between spaces (`$<<<$': p$<$.001, `$<<$': p$<$.01, `$<$': p$<$.05)}
  \label{tab:liwc_all_features_count_only}
\end{table}

We have shown that supporters change their linguistic styles between homogeneous and cross-cutting communication spaces. Now we ask whether users change their linguistic styles in actual interactions between like-minded and opposing users.
For this, we compare LIWC features by interaction types in each communication space.

Table~\ref{tab:liwc_all_features_count_only} counts the number of LIWC categories that show significant changes between interaction types and communication spaces.  
For example, when Trump supporters talk to Trump supporters, they show different linguistic behavior in 35 LIWC categories as compared to when they talk to Clinton supporters in \texttt{/r/The\_Donald}. 


\begin{table*}[th!]
  \footnotesize \frenchspacing
  \begin{tabular}{ll|cccccc|cccccc}
    \toprule
    & & \multicolumn{3}{c}{\tiny{$S_{C} \rightarrow S_{C}$ vs $S_{C} \rightarrow S_{T}$}} &  \multicolumn{3}{c|}{\tiny{${S_{C} \rightarrow S_{C}}$ vs ${S_{C} \rightarrow S_{U}}$}} & \multicolumn{3}{c}{\tiny{${S_{T} \rightarrow S_{T}}$ vs ${S_{T} \rightarrow S_{C}}$}}  &  \multicolumn{3}{c}{\tiny{${S_{T} \rightarrow S_{T}}$ vs ${S_{T} \rightarrow S_{U}}$}} \\
    \hline
    & & clinton & politics & news & clinton & politics & news & donald & politics & news & donald & politics & news \\
    \hline
    \hline
    \#words & $\uparrow$ & \cellcolor{red!25}\tiny{$>>>$} &  \cellcolor{blue!25}\tiny{${<<<}$} & \cellcolor{blue!25}\tiny{$<<<$} & \cellcolor{red!25}\tiny{$>>>$} & \cellcolor{blue!25}\textbf{\tiny{$<<<$}} & \cellcolor{blue!25}\textbf{\tiny{$<<<$}} & \cellcolor{red!25}\tiny{$>>>$} & \cellcolor{red!25}\tiny{$>>>$} & \cellcolor{red!25}\tiny{$>>>$} & \cellcolor{red!25}\tiny{$>>>$} & \cellcolor{blue!25}\textbf{\tiny{$<<<$}} & \cellcolor{blue!25}\textbf{\tiny{$<<<$}}\\
    \textbf{Category} & & & & & & &  & & & & & &  \\
    f. 1SG & $\uparrow$ & \cellcolor{red!25}\tiny{$>>>$} & \cellcolor{red!25}\tiny{$>>>$} & - & \cellcolor{red!25}\tiny{$>>>$} & - & - & \cellcolor{red!25}\tiny{$>>>$} & \cellcolor{red!25}\tiny{$>>>$} & \cellcolor{red!25}\tiny{$>>>$} & \cellcolor{red!25} \tiny{$>>>$} & - & -\\
    f. 1PL & $\uparrow$ & - & - & - & - & - & - & \cellcolor{blue!25}\tiny{$<<<$} & - & - & - & - & -\\
    f. 2. & $\uparrow$ & \cellcolor{blue!25}\tiny{$<<<$} & \cellcolor{blue!25}\tiny{$<<<$} & \cellcolor{blue!25}\tiny{$<<$} & \cellcolor{blue!25}\tiny{$<<<$} & \cellcolor{blue!25}\tiny{$<<$} & - & \cellcolor{blue!25}\tiny{$<<<$} & \cellcolor{blue!25}\tiny{$<<<$} & - & \cellcolor{blue!25}\tiny{$<<<$} & \cellcolor{blue!25}\tiny{$<<<$} & -\\
    f. positive & $\downarrow$ & - & - & - & \cellcolor{red!25}\tiny{$<<<$} & - & - & \cellcolor{red!25}\tiny{$<<<$} & - & - & \cellcolor{red!25}\tiny{$<<<$} & \cellcolor{blue!25}\tiny{$>>>$} & -\\
    f. negative & $\uparrow$ & - & - & - & - & - & - & - & - & - & \cellcolor{blue!25}\tiny{$<<$} & - & -\\
    f. ? & $\downarrow$ & \cellcolor{red!25}\tiny{$<<<$} & - & - & - & - & - & \cellcolor{red!25}\tiny{$<$} & \cellcolor{red!25}\tiny{$<<<$} & \cellcolor{red!25}\tiny{$<$} & \cellcolor{red!25}\tiny{$<<<$} & - & -\\
    
    \textbf{Score} & & & & & & & & & & & & & \\
    arousal & $\downarrow$ & \cellcolor{blue!25} \tiny{$>>$} & \cellcolor{blue!25} \tiny{$>$} & - & \cellcolor{blue!25} \tiny{$>>>$} & - & - & \cellcolor{blue!25} \tiny{$>>$} & - & - & \cellcolor{blue!25} \tiny{$>>$} & \cellcolor{red!25}\tiny{$<$} & -\\
    valence & $\downarrow$ & \cellcolor{blue!25}\tiny{$>>>$} & \cellcolor{blue!25}\tiny{$>>>$} & - & - & \cellcolor{blue!25}\tiny{$>>$} & - & - & \cellcolor{blue!25}\tiny{$>>>$} & \cellcolor{blue!25}\tiny{$>$} & \cellcolor{blue!25}\tiny{$>>>$} & \cellcolor{blue!25}\tiny{$>>>$} & \cellcolor{red!25} \tiny{$<<$}\\
    \hline
     & &+3/-3 & +4/-1 & +2/0 & +2/-3 & +3/0 & +1/0 & +3/-4 & +2/-3 & +1/-3 & +4/-4 & +4/-1 & +1/-1 \\
    \bottomrule
    
  \end{tabular}
  \caption{Linguistic features that pass a Bonferroni-corrected significance test. In all feature testing tables, the number of left/right angle bracket indicates the level of p-value, while the direction shows the relative relationship between positive instances and negative instances, `$<<<$': p$<$0.001, `$<<$': p$<$0.01, `$<$': p$<$0.05. We color cells with blue when supporters of one candidate are being more persuasive to supporters of opposite candidate; otherwise, we use red. We omit reporting full results due to space limitations.}
  \label{tab:liwc}
  \vspace{-1mm}
\end{table*}

Overall, the results align very well with our previous findings.
We observe that the first-person singular pronoun (1SG) is used more often ($p$ $<$ 0.001) in interactions among like-minded supporters than between supporters of opposing candidates. The use of first-person pronouns reflects speakers focusing attention onto themselves~\cite{tausczik2010psychological}. In politics, this is shown, for example, by positive political ads using more self-references than negative ads referencing opposing candidates~\cite{gunsch2000differential}. Similarly, we observe a greater use ($p$ $<$ 0.001) of the second-person pronoun (2) in interactions between supporters of opposing candidates than among like-minded users. This corresponds with repeated findings that the use of the second-person pronoun is negatively correlated with relationship quality~\cite{simmons2008hostile}. 
We also find that the use of question marks is greater in interactions between supporters of opposing candidates. This shows that asking/responding behavior is more common for these interactions.

Verb tense is a final LIWC category of interest. 
While personal pronouns show the subject's attention allocation, verb tense shows its temporal allocation~\cite{tausczik2010psychological}. We observe that Trump supporters use more verbs in the past tense ($p$ $<$ 0.01) in \texttt{/r/politics} when interacting with the supporters of Clinton than others who support Trump. This again corresponds with findings on negative political ads~\cite{gunsch2000differential}, which show that negative ads typically tackle past actions of the opponent.

Overall, we observe that linguistic patterns change in similar ways (e.g., the use of personal pronouns, etc.) when users interact in different communication spaces and when they interact with like-minded or opposing users.

\subsubsection{Do users follow a persuasive style when communicating with supporters of opposing candidates?}

We now examine whether interactions between supporters of opposing candidates correspond with a persuasive communication style or appear to be more expressive in nature.
As discussed in Section~\ref{sec:linguistic_features}, our analysis is based on LIWC and additional categories proposed by Tan et al. (\citeyear{tan2016winning}).
The up/downward arrows in the first column of Table~\ref{tab:liwc} show the correlation between each category and persuasive communication style. For example, a longer argument (\#words $\uparrow$) is likely to be more persuasive.  Additionally, we color each cell ``blue'' if supporters of either candidate employ a more persuasive style when communicating with supporters of the opposing candidate and ``red'' if they do so in a less persuasive style. 

Clinton supporters, when interacting with Trump supporters in the homogeneous space  (\texttt{/r/hillaryclinton}), 
use fewer words,
fewer first-person pronouns, 
more second-person pronouns, 
more question marks,
fewer high-arousal words (are thus calm), and
fewer high-valence words (are thus less happy). 
Three of these features correspond to persuasive style and three correspond to natural expression. 
%
When interacting with Trump supporters in the cross-cutting space (\texttt{/r/politics}), 
they use more words, 
fewer first-person pronouns, 
more second-person pronouns, 
fewer high-arousal words (are thus calm), and
fewer high-valence words (are thus less pleasant). 
Four of these features correspond to persuasive style, while only one corresponds to natural expression. 
In \texttt{/r/news}, two correspond to persuasive style while none correspond to natural expression. 

When Trump supporters talk to Clinton supporters in the homogeneous space  (\texttt{/r/The\_Donald}), they
use fewer words, 
fewer first-person singular pronouns, 
more first-person plural pronouns, 
more second-person pronouns, 
more positive words, 
more question marks, and
fewer high-arousal words (are therefore calm). 
Three of these features correspond to persuasive style, while four correspond to natural expression.
Trump supporters, when interacting with Clinton supporters in the cross-cutting space (\texttt{/r/politics}), 
use fewer words, 
fewer first-person singular pronouns, 
more second-person pronouns, 
more question marks, and
fewer high-valence words (are therefore less pleasant).
Two of these features correspond to persuasive style, while three correspond to natural expression.
In \texttt{/r/news}, one feature corresponds to persuasive style, while three correspond to natural expression. 

For the homogeneous spaces, (\texttt{/r/hillaryclinton} and \texttt{/r/The\_Donald}), the results are mixed. This indicates that supporters of both candidates follow neither style exclusively or even predominantly when interacting with each other. However, in the cross-cutting space (\texttt{/r/politics}), we observe that Clinton supporters follow more persuasive style than do Trump supporters.
For Clinton supporters, four of ten signals indicate they are engaging Trump supporters in a linguistic style akin to persuasive arguments. For Trump supporters, only two of ten signals point to them engaging Clinton supporters in an attempt at persuasive arguments.\footnote{We find similar patterns, but to a weaker degree, for interactions in \texttt{/r/news}, where political issues are less prevalent.}
In short, Clinton supporters appear to be using persuasive arguments in engaging supporters of opposing candidates, while Trump supporters do not. To understand these diverging tendencies better, we examine interactions with unassigned users.

In interactions with unassigned users, Clinton supporters use similar linguistic styles as when interacting with Trump supporters. But their styles change less compared with interactions between like-minded users. 
This indicates that Clinton supporters change their linguistic style more strongly when interacting with Trump supporters than when interacting with users of unidentified political leaning. 
Trump supporters exhibit quite different patterns, especially in cross-cutting spaces. When interacting with unassigned users in \texttt{/r/politics}, they tend to adopt a more persuasive style, using more words, more second-person pronouns, and fewer positive words as compared to interactions with Clinton supporters. 
In other words, although Trump supporters interacted with Clinton supporters less in linguistic styles indicative of persuasive argument, they did so much more strongly when interacting with users of unidentified political affiliation.
We find no significant difference when they communicate in the homogeneous space (\texttt{/r/The\_Donald}). 

We also find variances in the shift of linguistic styles by user types introduced in Section~\ref{sec:user_classification}.
Users who obtain high average scores in both homogeneous and cross-cutting spaces are more likely to vary their linguistic styles in engaging with supporters of the opposing candidate. 
We find this trend among both Clinton and Trump supporters. 
Sophistication and cultural knowledge of the communication space, as expressed by the high ratings in different environments, translate into higher adaptability for different communicative contexts. 
This corresponds with previous findings indicating that political engagement affects language use on Twitter~\cite{preoctiuc2017beyond}. 

In summary, we find significant linguistic style changes depending on whom users talk to and the communication space in which they talk. 
While findings differ between communication environments, we also observe consistent patterns of linguistic changes between interactions in cross-cutting spaces and interactions with supporters of opposing candidates. This may point to some inherent features of political communications across partisan lines. We leave a deeper exploration of this pattern for future work.
Overall, supporters of either candidate tend not to follow persuasive argumentation style in homogeneous communication spaces when talking with supporters of opposing candidates. By contrast, in cross-cutting spaces, Clinton supporters tend to employ the style of persuasive argument in talking with Trump supporters, while Trump supporters do not do the same when engaging Clinton supporters. However, Trump supporters adopt persuasive styles when talking with unassigned users.

\subsection{Wordings and Semantics}
\label{sec:common_words}

As do linguistic patterns, words and their semantic meaning also allow inferences on the characteristics of political communication. 
If supporters of opposing candidates use the same words prominently, we can infer that their political agendas are well aligned.
By contrast, a high variance between words used by both groups could point to a highly split view of which topics matter. 

\subsubsection{Common words}

\begin{figure} [ht!]
  \begin{center}
  \includegraphics[width=90mm]{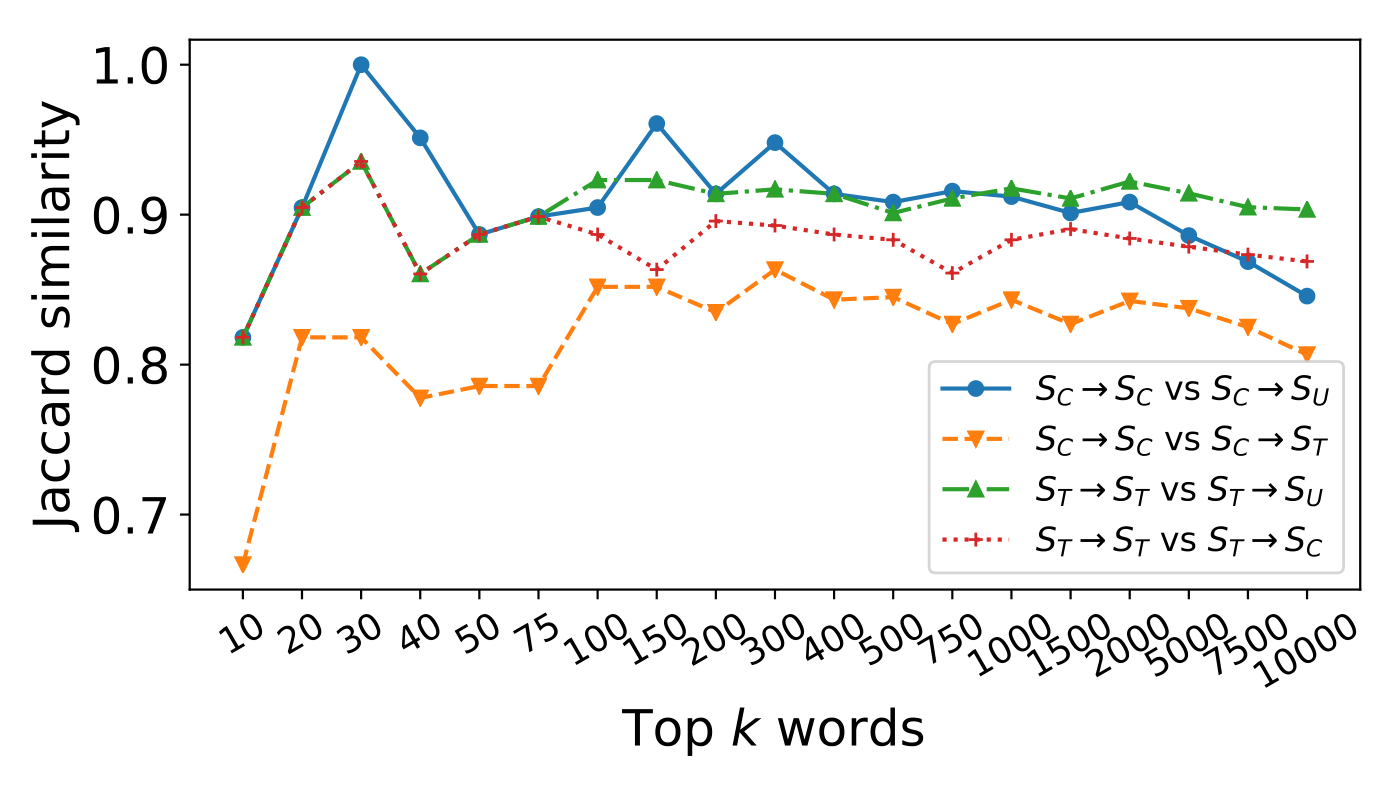}
  \caption{Jaccard similarity of the top $k$ words by TF-IDF used in two interaction types}
  \label{fig:common_words_politics}
  \end{center}
\end{figure}

Figure~\ref{fig:common_words_politics} shows the Jaccard similarity of words (the top $k$ words ranked by TF-IDF values) used in two interaction types by varying $k$. 
Comparing Clinton supporters interacting with other Clinton supporters and unassigned users, word use varies by 12.3\% on average.
When Clinton supporters interact with Trump supporters, the word difference increases to an average of 19.8\%. 
Trump supporters show similar patterns in differences between words used in different interaction types. 
Except for the first data points, which fluctuate strongly due to the comparatively few words that were considered, the Jaccard similarity becomes stable with growing $k$. 
In general, interactions between supporters of the same candidate or politically unassigned users share  do interactions with supporters of the opposing candidate.
This word choice gap exists both for Clinton and Trump supporters.\footnote{Due to space limitations, we report only the result for \texttt{/r/politics}. We observe similar patterns in \texttt{/r/news}.}

Although there are differences across interaction types in the words used by supporters, the overall similarities of word use are striking. 
This points to a shared agenda between supporters of different candidates dominated by the campaign for the U.S. election in 2016. 
This also raises an interesting challenge for further research: measuring the correlation between the similarity of word use between partisans and the salience of political events. 

We examine the top words commonly used across supporter groups and interaction types. To capture better the topics discussed, we focus on the top 30 nouns by TF-IDF. 
We observe common words across all four interaction types, mainly falling into two categories:
1) politicians: trump, clinton, hillary, bernie, sanders, obama; and
2) elections: president, win, vote, election, campaign, party.
These words are considered important in interactions of all four types. 
Given that our dataset covers the U.S. presidential election in 2016, it makes sense that these topics dominate among users regardless who they support or with whom they talk.
It is notable that a few words are more prominent in one interaction type than the other. For example, words such as ``media,'' ``news,'' and ``county'' are ranked only in interactions initiated by Trump supporters, whereas ``state'' and ``race'' are ranked only in interactions initiated by Clinton supporters.  
``Win'' is ranked in interactions among supporters who communicate among themselves. 
The results show that even simple TF-IDF can identify topical characteristics of different interaction types. 

\subsubsection{Word semantics}

While there are large commonalities with regard to the top words used by supporters of either candidate, the surrounding contexts of these words could vary significantly based on different interpretations of the same issue or event.
Domain-specific differences in the semantic meaning of words are a prominent research topic~\cite{hamilton2016diachronic,an2018semaxis}.
As we explained in Section~\ref{sec:linguistic_features}, we quantify semantic differences in the context of words used by supporters of opposing candidates. 

To illustrate this approach, let us examine the varying semantic contexts of the word ``women,'' which was one of the most mentioned topical words during the 2016 U.S. presidential election~\cite{benkler2017partisanship}. 
The cosine distance for the word ``women'' for Clinton supporters talking  among themselves versus talking to unassigned users is 0.24 ($S_{C} \rightarrow S_{C}$ vs. $S_{C} \rightarrow S_{U}$). For Trump supporters the cosine distance for the same word across the same interaction types is 0.16 ($S_{T} \rightarrow S_{T}$ vs. $S_{T} \rightarrow S_{U}$). 
By contrast, the semantic context of ``women'' changes significantly when supporters of opposing candidates interact, with a cosine distance of 0.35 for interactions initiated by Clinton supporters ($S_{C} \rightarrow S_{C}$ vs. $S_{C} \rightarrow S_{T}$) and 0.59 for interactions initiated by Trump supporters ($S_{T} \rightarrow S_{T}$ vs. $S_{T} \rightarrow S_{C}$). 


Examining the top 20 context words for ``women'' for each interaction type offers insight into what drives these differences. The context words for ``women'' among Clinton supporters relate to women (e.g., feminists, feminism) and minority issues (e.g., gays, minorities). When Clinton supporters interact with Trump supporters, the context words shift to feminism (e.g., feminists, genders) and sexual harassment (e.g., groping, assaults, unwanted, restroom). 
Among Trump supporters, context words point to fantasizing about women (e.g., fantasized, fantasizing, fantasizes). When interacting with Clinton supporters, Trump supporters use the term ``women'' in the context of homosexuality (e.g., lesbians, homosexually) and sexual harassment (e.g., accusers, harassed).
It is clear that both Clinton and Trump supporters use very different context words when they talk to supporters of the opposing candidate compared  to when talking among themselves.   
Using word embeddings, our approach is able to identify meaningful differences in the linguistic behaviors of users that remained hidden when simply focusing on the variety of words in use.

We now examine the degree of difference in word semantics of the same word across interaction types by using cosine distance. We focus on the top $k$ common words by varying $k$ from 10 to 10,000 across the interaction types.
Clinton supporters use words in almost the same context when they are interacting with other Clinton supporters and when they are interacting with unassigned users; on average, the difference of cosine distances is 0.01 up to the 2K top words mark. Beyond this point, word use begins to diverge. 
Trump supporters also use similar context words when interacting with other Trump supporters and when interacting with unassigned users. Their average difference of cosine distances is 0.05, slightly greater than that of Clinton supporters. Further, Trump supporters tend to use words in a different context when talking to Clinton supporters compared to when talking to unassigned users.  

We rank the common words (when $k$=300) by the cosine distance values to determine which words change their meaning the most when supporters of either candidate are talking to supporters of opposing candidate. We then examine the top 30 words with the highest semantic displacement values when talking with supporters of the opposing candidate.  
We find that words whose semantic meanings are changed dramatically are related to political issues, such as ``war,'' ``women,'' ``job,'' ``tax,'' ``law,'' ``healthcare,'' ``America,'' ``debate,'' ``government,'' ``fact,'' and ``history.'' This shows that the two supporter groups use different context words for these issues. Thus, while supporters share common issues, they think about them differently. 

In summary, Trump supporters use similar sets of words when talking to unassigned users and when talking to Clinton supporters. However, they change the meaning of words significantly depending on whom they talk to, expressed by differences in context words. 
Clinton supporters use fewer similar words but also shift the context of words across different interaction types. This reveals the complicated nature of political communication and shows the need for domain-specific linguistic analysis to capture the varying semantic meaning of words depending on their context.

\section{Discussion}

With regard to our first research question, whether supporters of different candidates interact in politically homogeneous and heterogeneous communication spaces, we can answer in the affirmative. The supporters of Clinton and Trump did interact directly in the subreddits \texttt{/r/politics} and \texttt{/r/news}. This was true for 77.9\% of all users active in \texttt{/r/hillaryclinton}, and for 60.9\% of all users active in \texttt{/r/The\_Donald}. 
Thus, while not every user active in politically homogeneous communication spaces chose to engage actively in heterogeneous communication environments, for both candidates a majority of the supporters did so.
Still, only slightly more than a third of threads in these heterogeneous spaces contained direct interactions between Clinton and Trump supporters. So, while cross-cutting interactions between partisans took place in heterogeneous communication environments, they were far from the norm.

We observed that communication styles change depending on where and with whom supporters talk. 
The political composition of a communication space affected the way a candidate's supporters interacted with other users holding the same opinions.  We found interactions between supporters of the same candidate to feel more open and comfortable and to exhibit greater group responsibility in homogeneous environments than in the same type of interactions in heterogeneous environments. 
We also found that supporters changed their communication styles in interacting with supporters of opposing candidates compared to interacting among their fellow supporters. Linguistic styles varied more for the interactions in homogeneous spaces compared with interactions in cross-cutting spaces. Trump supporters had more significant changes than Clinton supporters. However, the findings show that Clinton supporters were more likely to shift into persuasive argumentation style when interacting with Trump supporters than vice versa. 
%
These differences were most pronounced for highly active users whose contributions were, on average, highly scored by other Reddit users. Conscious adjustment of one's communicative strategy to differently structured communication spaces thus seems to go hand in hand with sophistication and cultural knowledge of the communication space Reddit. While supporters of different candidates used similar words in interactions, we found the semantic context of these words to vary clearly between partisans of different stripes. 

These findings indicate that both the pessimistic expectations of ``echo chambers'' and the optimistic view of online spaces enabling political discourse are in part correct in representing Reddit as a political communication space. Political talk on Reddit corresponds with elements of both views, but not completely with either view. 
We find that only a minority of users active in politically homogeneous communication limit their participation exclusively to such environments. 
This might be taken by some as support for an ``echo chamber'' view of political talk on Reddit, at least for this minority. 
But we also find many users active in these homogeneous spaces to engage actively in cross-cutting exchanges with supporters of opposing candidates in heterogeneous communication environments.
To a significant degree these users also contextually adapt their communication strategies, expressed by shifts in the linguistic structure of their posts. 
This might be taken as an indicator of Reddit as an environment for discursive exchanges across partisan lines. Yet these exchanges did not necessarily follow a persuasive style of argument, deemed vital in productive exchanges between political opponents. This raises the possibility that while some exchanges may actually conform to the hopes of democratic theorists, others might follow less constructive communication strategies, such as, for example, grand standing, confrontation, or downright incivility.

It is interesting to note the differences in the way supporters of Hillary Clinton and Donald Trump behaved. Although Trump supporters dwarf Clinton supporters in raw numbers both in homogeneous and heterogeneous communication environments, they are, relatively speaking, much more comfortable interacting among themselves than are Clinton supporters. Even if they expose themselves to cross-cutting political talk, they do so much less actively than Clinton supporters.

In combination, our findings point to a complicated picture of online political discourse. 
The current academic debate tends to treat online communication spaces as ``echo chambers''--spaces in which like-minded people interact, lose sight of other social and political groups, and thus over time become more extreme in their opinions leading to a polarization of political discourse ~\cite{Sunstein:2017aa}.
Although, as discussed, this view is empirically highly contested, it still serves as a baseline in discussing political discourse online. This leads researchers to search for evidence for or against ``echo chambers'' rather than examining actual interaction patterns in political discourse online. 
This is problematic because the concept of ``echo chambers'' combines structural patterns--interactions in homogeneous communication environments--deterministically with outcomes--loss of information about and empathy for the other side. 
Here, we feel researchers would be better served using more precise conceptualizations in their work that would allow for viewing structural patterns and outcomes independently rather than as part of some conceptual package, such as the ``echo chamber.''
We have shown that the analysis of user behavior in politically homogeneous and cross-cutting communication environments allows for a differentiated account of behavior and behavioral shifts depending on the communicative context within which people interact. 
The picture that emerges points to political discourse online as a complicated mix of different behaviors across contexts. To do justice to this complicated reality we should dispense with simplistic arguments and instead employ more flexible concepts to capture the complicated reality of interconnections and inter-dependencies in political discourse in online spaces.

Our findings also demonstrate the importance of accounting for the nature of communication spaces and interaction types in understanding political communication.  While previous work reveals the linguistic differences of liberals and conservatives on Twitter~\cite{sylwester2015twitter}, our work shows that such linguistic features can vary between homogeneous and cross-cutting communication spaces. This varying linguistic signature should be considered in broader research efforts aimed at predicting gender, age, preference, or other personality characteristics based on linguistic features.

While this study adds to our understanding of political communication online in homogeneous and cross-cutting spaces, it is not without limitations. First, the Reddit data we used might be incomplete.  Gaffney and Matias (\citeyear{gaffney2018caveat}) reported recently that 0.043\% of comments and 0.65\% of submissions may be missing in widely used Reddit datasets. This also concerns the dataset used here. Yet, considering recent efforts that successfully replicated previous studies with newly crawled data\footnote{http://www.cs.cornell.edu/~jhessel/reddit/gaps.html} and further discussion by Gaffney and Matias (\citeyear{gaffney2018caveat}), we conclude that for our analyses the effect of missing comments would be marginal. Second, as our work is based on the supporter groups of Clinton and Trump during the 2016 U.S. presidential election, group biases may be more noticeable than in general political discussions between people with different political leaning in less hotly contested elections or between elections. Comparative longitudinal studies covering various political events in various countries are required to allow for the generalization of our findings.

\section{Conclusion}

In summary, our paper makes three contributions. First, we provide evidence that supporters of opposing candidates are active in politically heterogeneous environments and engage in ideologically cross-cutting political talk online. Second, we find that they adjust their linguistic style once they engage with political opponents in cross-cutting communication environments. 
Third, our observations suggest that the communication space we examined is inadequately covered by the ``echo chamber'' concept. We find Reddit to be a space, in which politically cross-cutting and homogeneous spaces and interactions co-exist.

  \bibliographystyle{aaai}
  \bibliography{FULL-AnJ.116}

\end{document}